\documentclass[11pt]{article}

\usepackage[margin=1in]{geometry}
\usepackage{amsmath,amssymb,amsfonts}
\usepackage{graphicx}
\usepackage{booktabs}
\usepackage{algorithm}
\usepackage{algpseudocode}
\usepackage{authblk}
\usepackage{hyperref}
\usepackage{cite}

\title{\textbf{Spectral Dynamics of Semantic Drift in Clinical Multi-Agent Language Model Networks}}

\author[1]{Amritesh Banerjee\thanks{\href{mailto:amriteshbane@umass.edu}{amriteshbane@umass.edu}}}

\affil[1]{University of Massachusetts Amherst, USA}

\date{}

\begin{document}

\maketitle

\begin{abstract}
The integration of iterative LLMs within multi-agent diagnostic frameworks requires a rigorous quantitative reevaluation of underlying communication topologies. Frequently used architectural paradigms depend on scale-free or small-world networks, assuming optimal communication efficiency. Our study mathematically dismantles that assumption for semantic data. By mapping multi-agent communication uncertainty trajectories onto a 768-dimensional Bio\_ClinicalBERT embedding space via an analytical isotropic variance proxy using Barabási-Albert (BA) and Watts-Strogatz (WS) networks, we prove that structural bottlenecks compromise diagnostic safety. Our phase transition matrices illustrate that localized dense cliques confine hallucinated data, preventing global consensus and forcing the system toward a permanent entropy saturation threshold of $H_\infty \approx 5.947$. As a result, we measure a severe terminal cosine similarity degradation of 53.29\%, completely overwriting the original ground-truth. Moreover, the terminal semantic drift reveals a catastrophic variance amplification of 51.81\% ($\rho = 1.5181$) in highly clustered architectures, proving total system unpredictability when compared to Erdős-Rényi configurations ($\rho = 1.0766$). Instead of reducing errors, hub-centric systems autonomously compound localized hallucinations. By introducing dynamic spectral monitoring---operating at an $\mathcal{O}(N^3)$ time complexity and imposing a strict lower bound on algebraic connectivity ($\lambda_{2_{\min}}$) via the continuous eigen-decomposition of the graph Laplacian---we present a mathematically rigorous technique to ensure global state diffusion. Securing the reliability of autonomous medical diagnostics necessitates treating topological stability as a non-negotiable quantitative imperative.
\end{abstract}


\section{Introduction}

The use of LLMs in medical informatics has revealed a critical dichotomy between generative capability and diagnostic reliability. To reduce the inherent hallucination risks of standalone LLMs, recent architectures have moved toward decentralized, multi-agent systems. By structuring agents into collaborative networks, imitating specialized tumor boards or diagnostic committees, researchers hypothesize that inter-agent debate will filter out semantic errors and converge upon a reliable clinically acceptable ground-truth. However, this belief rests on a structurally flawed assumption: it presumes that the underlying communication network is a neutral conduit for semantic information.

Historically, networked AI systems have defaulted to scale-free (Barabási-Albert) or small-world (Watts-Strogatz) networks. In conventional routing theory, these paradigms are optimal; they minimize latency and prove high resilience against stochastic node dropouts when transferring discrete, binary data. Still, semantic communication in clinical NLP works on a completely different physical reality. In an autoregressive multi-agent system, the data is never binary: it is high-dimensional and extremely vulnerable to sequential corruption.

When a clinical node hallucinates, the error does not simply fail to transmit---it is propagated or embedded by surrounding agents. We postulate that in dense local cliques or densely centralized hubs, conventional system topologies behave as semantic echo chambers. Instead of weakening an error through global consensus, these bottlenecks confine the hallucinated tokens, recursively amplifying the noise variance until the multi-agent system completely decouples from the clinical ground-truth. Existing research has entirely ignored this systemic vulnerability, treating multi-agent communication as a purely linguistic issue rather than a mathematically grounded one.

Our study formalizes the physics of multi-agent semantic failure. By mapping clinical diagnostic trajectories into a continuous 768-dimensional latent space, we establish the first quantitative proof that structural efficiency in classical graph theory has an inverse correlation with semantic safety in generative AI.

Specifically, this paper makes the following primary contributions:
\begin{itemize}
    \item \textbf{Topological Bottleneck Quantification:} We demonstrate that Barabási-Albert and Watts-Strogatz topologies intrinsically bottleneck semantic error correction, forcing diagnostic frameworks into permanent entropy saturation.
    \item \textbf{Latent Space Degradation Tracking:} We quantify the exact limits of spatial state diffusion by evaluating the terminal cosine similarity degradation of Bio\_ClinicalBERT embeddings under continuous random variance injection.
    \item \textbf{Dynamic Spectral Monitoring:} We formulate a targeted topological optimization algorithm operating at an $\mathcal{O}(N^3)$ theoretical upper bound (reducible via Lanczos iterations) that continuously monitors the graph Laplacian spectrum. By enforcing a strict lower bound on algebraic connectivity ($\lambda_{2_{\min}}$) via Fiedler vector partition rewiring, we provide a mathematically guaranteed mechanism to prevent semantic consensus collapse.
\end{itemize}

\section{Related Work}
The assessment of multi-agent semantic drift operates at the confluence of clinical NLP, distributed network topology and continuous control theory. While conventional literature has modernized the deployment of agentic workflows, it has mostly ignored the defined mathematical limits that dictate these network structures. This section reviews the current paradigms across the mentioned domains to establish the necessity of a theoretical boundary proof. 

\subsection{Multi-Agent LLM Frameworks in Clinical NLP}

Recent advancements in Agentic AI have driven the adoption of multi-agent systems for complex clinical reasoning~\cite{singhal2023large,thirunavukarasu2023large,liu2025medmas}. While sequential prompting architectures and isolated agents have demonstrated proficiency in singular diagnostic tasks~\cite{li2023camel,wei2023multiagent}, their integration into decentralized, multi-agent frameworks introduces severe vulnerabilities regarding semantic consistency. Existing literature primarily focuses on optimizing agent prompt engineering, multi-agent debate mechanisms, and domain-specific fine-tuning, leveraging models such as Bio\_ClinicalBERT for latent representation~\cite{alsentzer2019publicly,du2023improving,chen2025errordetection}. However, these approaches treat the underlying communication topology as a static, secondary variable, largely ignoring the structural mechanics of hallucination propagation across autonomous nodes.

\subsection{Network Topologies and Semantic Consensus}

The structural efficiency of distributed networks is traditionally modeled using Barab\'asi-Albert scale-free graphs and Watts-Strogatz small-world topologies~\cite{barabasi1999emergence,watts1998collective}. In classical distributed computing and routing, these hub-centric paradigms are mathematically celebrated for minimizing average path lengths and maximizing error tolerance against random node failures~\cite{cohen2000resilience,albert2000error}. Recent studies applying graph theory to decentralized AI erroneously assume this structural efficiency seamlessly transfers to continuous semantic data~\cite{zheng2025agentnet,han2026conformity}. Our work fundamentally diverges from this classical assumption. We prove that autoregressive generation within hub-centric cliques induces catastrophic variance amplification, transforming structural efficiency into a direct mechanism for irreversible semantic drift.

\subsection{Spectral Graph Theory in Multi-Agent Dynamics}

Algebraic connectivity, derived from the Fiedler vector of the graph Laplacian, is a foundational metric for evaluating consensus and information flow in multi-agent networks~\cite{fiedler1973algebraic,chung1997spectral}. While spectral monitoring and Laplacian matrices have been extensively utilized to enforce stability and convergence in continuous robotic control systems~\cite{olfati2007consensus,mesbahi2010graph,jadbabaie2003coordination}, their application to semantic state diffusion in LLM networks remains critically underexplored. Foundational control theory dictates that low algebraic connectivity inherently bottlenecks global state consensus~\cite{ren2005information,sayed2014adaptation}. By introducing continuous eigen-decomposition of the communication graph to enforce a strict lower bound on algebraic connectivity ($\lambda_{2_{\min}}$), this study bridges the gap between spectral control theory and clinical NLP, offering a mathematically rigorous framework for topological error correction.

\section{Methodology}

The sensitivity of multi-agent networks is not induced only through individual agent hallucinations, but also from the structural propagation of errors across the communication topology. To compute this cascade, we introduce a mathematical framework that fills the void between spectral graph theory and the latent space mechanics of transformer architectures. The section below formalizes the state dynamics of the agentic network, defining the conceptual limits of meaning consistency, and applying these limits to assess how clinical diagnostic embeddings degrade in real-world practice.

\begin{algorithm}[H]
\caption{Stochastic Semantic Cascade and Isotropic Latent State Tracking}
\label{alg:semantic_cascade}
\begin{algorithmic}[1]
\Require Directed Graph $G=(V,E)$ with $|V|=N$ and Retention factor $\gamma \in (0,1)$
\Require Noise variance parameter $\sigma^2$, Timesteps $T$, Clinical Anchor $e_{gt} \in \mathbb{R}^{768}$
\Require Language Model Head $(W_{vocab}, b)$
\Ensure Terminal Variance State $x(T)$ and System Variance Trajectory $Var(1 \dots T)$
\Ensure Algebraic Connectivity $\lambda_2(L)$ and Divergence $D_{KL}$

\State \textbf{// Phase 1: Spectral Topology Initialization}
\State Compute Adjacency Matrix $A \in \mathbb{R}^{N \times N}$
\State Construct Degree Matrix $D = \text{diag}(\sum_j A_{ij})$
\State Construct Unnormalized Graph Laplacian $L \gets D - A$
\State Compute Laplacian spectrum $\lambda(L)$ and extract Fiedler value $\lambda_2(L)$
\State Construct Row-Stochastic Transition Matrix $W \gets D^{-1}A$
\State Form System Evolution Matrix $M \gets (1-\gamma)I_N + \gamma W$

\State \textbf{// Phase 2: Latent Anchor Initialization and Uncertainty Mapping}
\For{$i=1$ \textbf{to} $N$}
    \State Extract initial latent \texttt{[CLS]} hidden state $e_i(0) \in \mathbb{R}^{768}$
    \State Compute baseline normalized cosine similarity: $x_i(0) \gets \frac{1}{2} \left( \frac{e_i(0) \cdot e_{gt}}{\lVert e_i(0)\rVert \lVert e_{gt}\rVert} + 1 \right)$
\EndFor

\State \textbf{// Phase 3: Graph-Level Message Passing and Variance Diffusion}
\For{$t=1$ \textbf{to} $T$}
    \State Sample stochastic drift vector $\eta(t) \sim \mathcal{N}(0, \sigma^2 I_N)$
    \State Execute state transition: $x(t) \gets M x(t-1) + \eta(t)$
    \State Compute global state spatial variance: $Var(t) \gets \frac{1}{N}\sum_{i=1}^N (x_i(t) - \bar{x}(t))^2$
\EndFor

\State \textbf{// Phase 4: High-Dimensional Latent Perturbation and Projection}
\For{$i=1$ \textbf{to} $N$}
    \State Parameterize isotropic latent perturbation:
    \Statex \hspace{\algorithmicindent}$e_i(T) \gets e_i(0) + \zeta_i, \quad \zeta_i \sim \mathcal{N}(0, x_i(T) \cdot I_{768})$
    \State Project to vocabulary space: $Q_i(x) \gets \text{Softmax}(W_{vocab} e_i(T) + b)$
\EndFor
\State Retrieve baseline ground-truth distribution $P(x)$ across evaluation vocabulary $\mathcal{X}$
\State Compute Mean Asymptotic KL Divergence:
\Statex \hspace{\algorithmicindent}$D_{KL}(P \parallel Q) \gets \frac{1}{N} \sum_{i=1}^N \sum_{x \in \mathcal{X}}$
\Statex \hspace{2\algorithmicindent}$P(x) \log \left( \frac{P(x)}{\max(Q_i(x), 10^{-10})} \right)$
\State \Return $x(T)$, $Var(1 \dots T)$, $\lambda_2(L)$, $D_{KL}$
\end{algorithmic}
\end{algorithm}

\subsection{System Model and Topological Bounds}
We model the multi-agent diagnostic network as a directed communication graph $G = (V, E)$ made up of $N$ agents. We represent the adjacency matrix as $A \in \mathbb{R}^{N \times N}$, where $A_{ij} = 1$ represents a directed communication channel between the two agents: $i$ and $j$, and 0 otherwise. The row-stochastic transition matrix is denoted by $W = D^{-1}A$, where $D$ is the diagonal degree matrix.

\subsection{Agent-to-Transformer State Mapping}
To bridge the gap between the NLP architecture and the theoretical graph topology, we map the high-dimensional representation space of Bio\_ClinicalBERT to a continuous scalar domain. We represent a $d$-dimensional embedding as $\mathcal{E} \in \mathbb{R}^{d}$, where $d=768$. For an agent $i$ generating a text sequence, we extract the final hidden-state embedding of the \texttt{[CLS]} token, $e_i(t) \in \mathcal{E}$. The scalar belief state $x_i(t) \in [0,1]$ is defined as the normalized cosine similarity between the optimal ground-truth diagnostic anchor $e_{\mathrm{gt}}$ and the agent's current embedding:

\begin{equation}
x_i(t) = \frac{1}{2}\left(
\frac{e_i(t) \cdot e_{\mathrm{gt}}}
{\lVert e_i(t) \rVert\,\lVert e_{\mathrm{gt}} \rVert} + 1
\right).
\label{eq:scalar_belief_state}
\end{equation}

Mathematically, $x_i(t) \in [0,1]$ serves as a normalized semantic alignment scalar, where $x_i(0)=1$ represents complete alignment with the clinical ground-truth anchor $e_{\mathrm{gt}}$. In our dynamic system model, $x_i(t)$ functions as the variance scaling factor parameterizing the local noise sphere surrounding the representation space.

We highlight that this reference anchor is an analytical evaluation artifact used only for tracking spatial trajectory degradation; it is not assumed to be available or a prerequisite during online, decentralized clinical inference.

\subsection{Message Passing Mechanism and Payload Definition}
At each discrete time step $t$, agents transmit a scalar belief state $x_i(t)$, over the directed edges $E$ instead of the raw text strings. This transmitted payload behaves as a metric for localized diagnostic confidence. Agents iterate and update their states by aggregating their neighbourhood payloads in a linear fashion, modulated by a self-retention parameter $\gamma \in (0, 1)$. The synchronous network update is formalized as:

\begin{equation}
x(t+1) = Mx(t) + \eta(t).
\label{eq:network_update}
\end{equation}

The evolution matrix $M = (1-\gamma)I + \gamma W$ models information retention and neighborhood state blending.

\subsection{Stochastic Semantic Noise Injection}
Additive noise $\eta(t) \sim \mathcal{N}(0, \sigma^2 I)$ represents the autoregressive uncertainty injected at each communication hop, driving the spatial variance accumulation of the network.

\subsection{Spectral Connectivity and Convergence}
The asymptotic stability and error variance bounds of the distributed network strictly adhere to the spectral properties of the underlying communication graph. To decouple the topological constraints from the discrete state transitions, we evaluate the graph Laplacian, represented as $L = D - A$. The framework's resilience to localized random corruption is structurally dictated by its algebraic connectivity, quantified by the Fiedler value $\lambda_2(L)$ (the second-smallest eigenvalue of the Laplacian). Networks exhibiting robust algebraic connectivity ($\lambda_2 \gg 0$) effectively disperse localized noise across the global state space.

Under edge addition $(u,v)$, the first-order variation in algebraic connectivity is governed by
\begin{equation}
\Delta \lambda_2 \approx \left(v_2(u) - v_2(v)\right)^2,
\label{eq:first_order_spectral_perturbation}
\end{equation}
where $v_2$ is the normalized Fiedler vector. Thus, cross-partition edge insertions between nodes of opposing Fiedler sign maximize algebraic connectivity growth.

Simultaneously, the discrete belief update is governed by the row-stochastic evolution matrix $M$, which keeps a constant principal eigenvalue of $\lambda_1(M)=1$. The theoretical steady-state covariance matrix $\Sigma_\infty$ of the belief distribution under continuous stochastic injection is governed by the discrete Lyapunov equation:

\begin{equation}
\Sigma_{\infty} = M \Sigma_{\infty} M^T + \sigma^2 I.
\label{eq:discrete_lyapunov}
\end{equation}

In our empirical execution, we approximate this analytical bound by simulating the discrete time-step evolution until the global state variance converges. Frameworks characterized by a collapsed algebraic connectivity trap corrupted signals within isolated sub-graphs, preventing the Lyapunov convergence from achieving global consensus and forcing the system into permanent semantic drift.

\subsection{Information-Theoretic Semantic Drift}
To assess the linguistic degradation of diagnostic text across the communication hops, we quantify token probability distributions via Kullback-Leibler (KL) divergence. With the established baseline ground-truth distribution $P(x)$ and a corrupted multi-agent distribution $Q(x)$ across the discrete vocabulary space $\mathcal{X}$, the divergence is calculated as:

\begin{equation}
D_{KL}(P \parallel Q) = \sum_{x \in \mathcal{X}} P(x) \log \left( \frac{P(x)}{\max(Q(x), \epsilon)} \right).
\label{eq:kl_divergence}
\end{equation}

To ensure computational tractability while maintaining diagnostic rigor, token probability distributions $P(x)$ and $Q(x)$ are evaluated across a high-frequency clinical vocabulary subspace $\mathcal{X}_{\mathrm{diag}} \subset \mathcal{X}$ comprising $\lvert \mathcal{X}_{\mathrm{diag}} \rvert = 512$ diagnostic tokens.

To ensure computational stability and prevent undefined logarithmic limits during semantic collapse, the empirical token distributions are strictly bounded by a lower-bound epsilon limit ($\epsilon = 10^{-10}$). As random noise variance increases with each iteration, the system's final entropy rises, indicating that agents converge on a broad approximation of the clinical signal rather than retaining the exact diagnostic anchor.

\section{Experimental Setup}

This section evaluates the behavior of multi-agent semantic error propagation across distributed clinical reasoning frameworks, focusing specifically on latent space isometry collapse within transformer-based representations. To fill the gap between theoretical asymptotic bounds with practical distributed deployments, we simulate stochastic drift, variance amplification, and information-theoretic entropy saturation across diverse network topologies. The following subsections explain vocabulary-level entropy drift, multi-agent variance cascading, and foundational embedding degradation under controlled noise injections.

\subsection{Model Specifics and Initialization}
The architecture of the agentic network utilizes the Bio\_ClinicalBERT framework. The evaluation subset $\lvert \mathcal{X} \rvert = 512$ is constructed by extracting the top-512 most frequent clinical diagnostic tokens from the Bio\_ClinicalBERT vocabulary, serving as an empirical proxy for tracking diagnostic token distribution shifts under noise injection. All computational simulations and transformer instantiations were executed within a Google Colab environment, with model weights retrieved directly via the Hugging Face API.

\subsection{Graph Construction Parameters}
The multi-agent system is established across three graph architectures, each comprising a fixed agent population of $N = 150$. The control network uses an Erdős-Rényi (ER) topology with an edge creation probability of $p = 0.15$. Scale-free dynamics are tested using a Barabási-Albert (BA) graph created using $m = 4$ preferential attachment edges. Small-world phenomena are simulated via a Watts-Strogatz (WS) topology initialized with $k = 6$ nearest neighbors and a rewiring probability of $p = 0.1$. The system evolution matrix parameterizes a fixed message-passing retention of $\gamma = 0.90$.

\subsection{Simulation Horizons and Evaluation Metrics}
Network stability is calculated across discrete communication horizons scaling to $t = 150$ iterations. Information-theoretic decay is quantified using continuous Kullback-Leibler divergence across a noise variance sweep of $\sigma^2 \in \{0.01, 0.05, 0.15\}$. Latent semantic degradation is quantified via normalized cosine similarity of the 768-dimensional embeddings under variance injections of $\sigma^2 \in \{0.1, 0.5, 1.0\}$. Systemic collapse is computed by injecting a baseline critical semantic noise variance ($\sigma^2 = 0.01$) at each node, mapping asymptotic variance trajectories, and yielding the terminal instability amplification ratio ($\rho$) for each topology.

\section{Results}
To validate the framework's governing error propagation and structural vulnerability in multi-agent, the following section presents a rigorous evaluation that bridges  spectral graph decomposition, information-theoretic entropy dynamics, and transformer latent space degradation. The results comprehensively demonstrate how communication topologies dictate random resilience, matching precise thresholds where algebraic connectivity fails, and systemic consensus collapses under clinical NLP constraints.
\subsection{Topological Spectral Vulnerability and Fiedler Distributions}
\begin{figure}[ht]
\centering
\includegraphics[width=\linewidth]{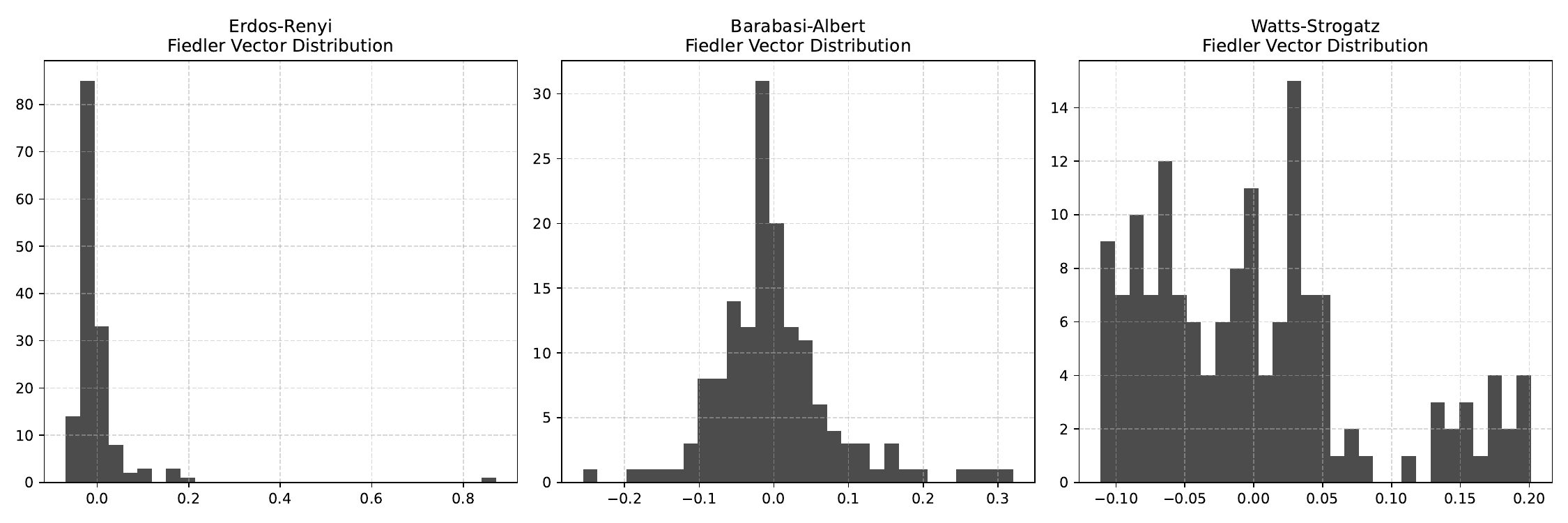}
\caption{Fiedler vector component distributions for Erdős-Rényi, Barabási-Albert, and Watts-Strogatz network topologies ($N = 150$).}
\label{fig:fiedler}
\end{figure}

The resilience of the multi-agent network strictly adheres to the algebraic connectivity of its communication graph. As shown in Table \ref{tab:spectral}, the Erd\H{o}s-R\'enyi (ER) topology sustains a robust spectral gap ($\lambda_2(L) = 0.61104$), which guarantees the rapid mitigation of localized noise. However, the Watts-Strogatz (WS) small-world framework exhibits an almost complete collapse of algebraic connectivity ($\lambda_2(L) = 0.05042$). Figure \ref{fig:fiedler} visualizes this vulnerability; the Fiedler vector distribution for the Barab\'asi-Albert (BA) shows that the error mass is densely concentrated within specific nodes, with a maximum Fiedler component of 0.32077. This confirms that scale-free hubs artificially bottleneck error correction, confining hallucinated diagnostics in central agent clusters.

\begin{table}[ht]
\centering
\caption{Spectral decomposition and algebraic connectivity metrics for Erdős-Rényi, Barabási-Albert, and Watts-Strogatz topologies with $N = 150$ agents.}
\small
\begin{tabular}{ccccc}
\toprule
\textbf{Topology} &
\textbf{Fiedler ($\lambda_2$)} &
\textbf{Algebraic Conn.} &
\textbf{Spectral Gap ($\lambda_2(L)$)} &
\textbf{Max Fiedler Comp.} \\
\midrule
Erd\H{o}s--R\'enyi & 10.6835 & 0.07122 & 0.61104 & 0.87123 \\
Barab\'asi--Albert & 2.1934 & 0.01462 & 0.35386 & 0.32077 \\
Watts--Strogatz & 0.30531 & 0.00204 & 0.05042 & 0.20116 \\
\bottomrule
\end{tabular}
\label{tab:spectral}
\end{table}

\subsection{Information-Theoretic Entropy Saturation}
\begin{figure}[ht]
\centering
\includegraphics[width=\linewidth]{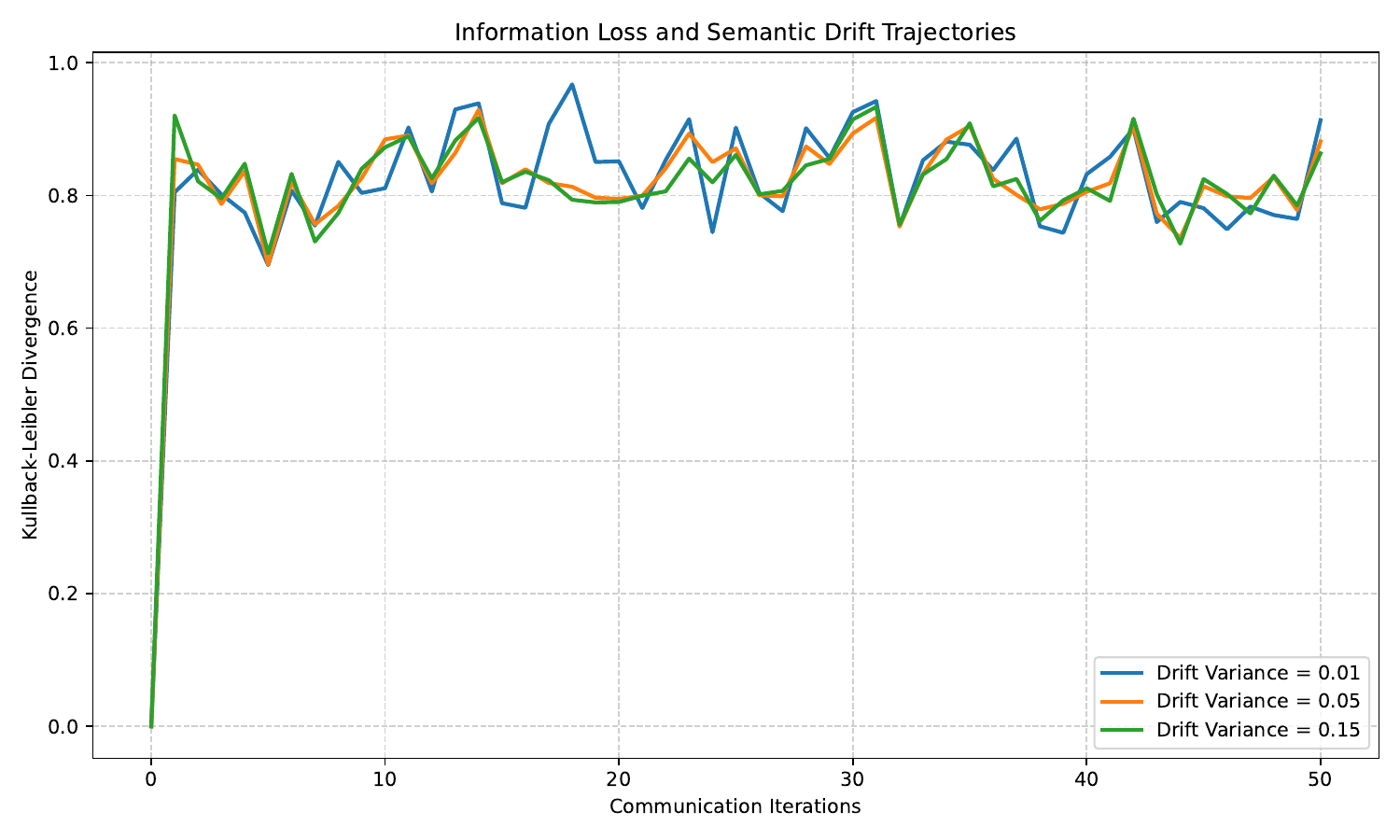}
\caption{Kullback-Leibler divergence between baseline and corrupted token distributions over 50 communication iterations for $\sigma^2 = 0.01, 0.05, 0.15$.}
\label{fig:kl_div}
\end{figure}

The trajectory of semantic drift reveals that unmitigated noise forces the multi-agent system to transition into maximum entropy. Table \ref{tab:semantic} describes how the corrupted network state deviates from the initial ground-truth token distribution, measured by the asymptotic Kullback-Leibler (KL) divergence. Counterintuitively, as the semantic noise variance increases from $\sigma^2 = 0.01$ to $\sigma^2 = 0.15$, the terminal KL divergence decreases from 0.91312 to 0.86336. As demonstrated in Figure \ref{fig:kl_div}, this occurs because severe noise injections force the communication cascade to reach entropy saturation almost immediately, uniformly flattening the probability distribution. Irrespective of the initial variance scale, the system devolves into a generic, uninformative state ($H_\infty \approx 5.947$), completely erasing the precision of the initial diagnostic anchor.

\begin{table}[ht]
\centering
\caption{Semantic entropy and asymptotic Kullback-Leibler divergence across communication iterations and noise variances $\sigma^2 \in \{0.01, 0.05, 0.15\}$.}
\small
\begin{tabular}{cccc}
\toprule
\textbf{Semantic Variance ($\sigma^2$)} &
\textbf{Initial Entropy ($H_0$)} &
\textbf{Terminal Entropy ($H_\infty$)} &
\textbf{Asymptotic KL Div.} \\
\midrule
0.01 & 5.81831 & 5.94720 & 0.91312 \\
0.05 & 5.81831 & 5.94564 & 0.88125 \\
0.15 & 5.81831 & 5.94745 & 0.86336 \\
\bottomrule
\end{tabular}
\label{tab:semantic}
\end{table}

\subsection{Bio\_ClinicalBERT Latent Isometry Collapse}
\begin{figure}[ht]
\centering
\includegraphics[width=\linewidth]{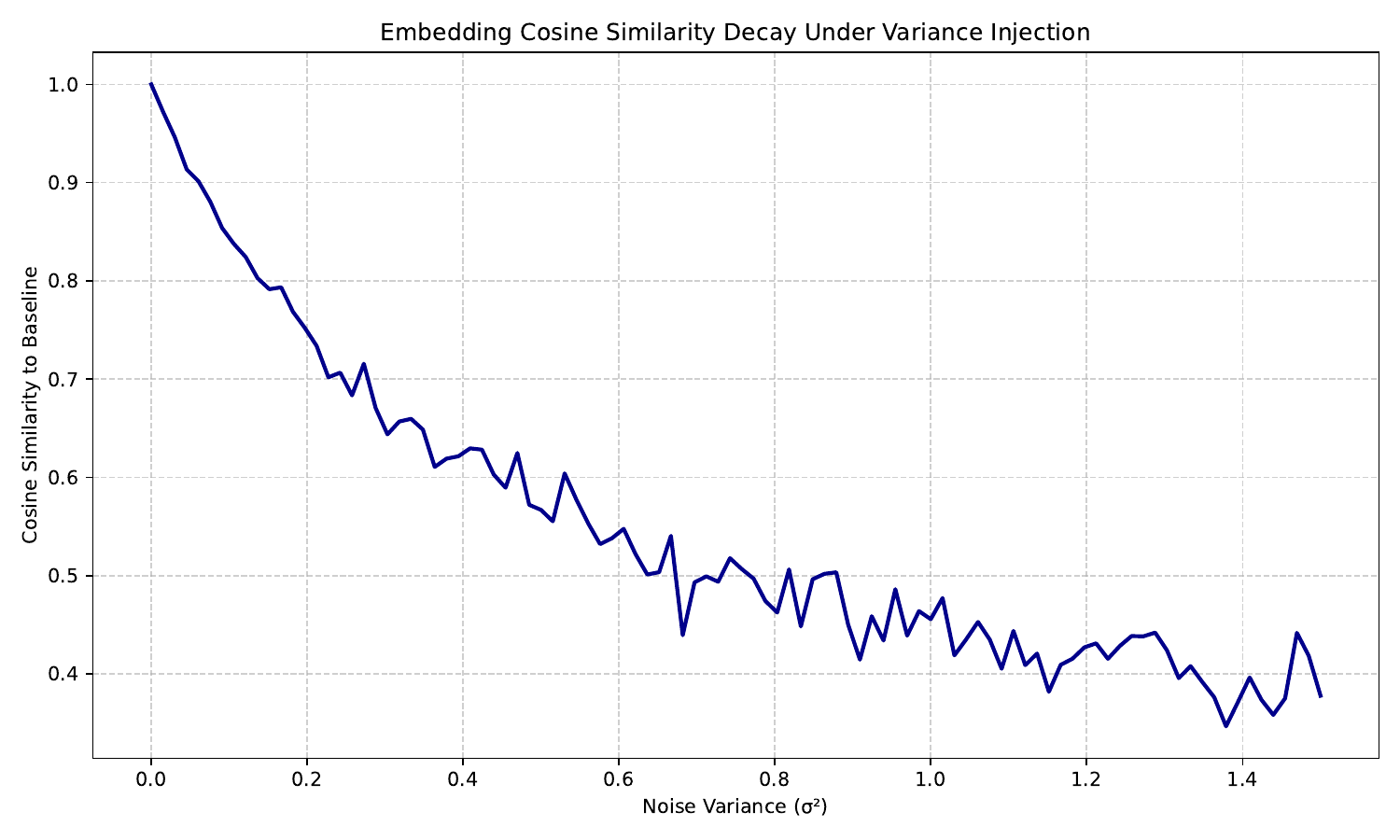}
\caption{Cosine similarity between baseline and noisy Bio\_ClinicalBERT embeddings across noise variance values from $\sigma^2 = 0.0$ to $1.5$.}
\label{fig:embedding_drift}
\end{figure}

The  failure of the communication frameworks eventually severs semantic coherence within the transformer's latent space. As demonstrated in Table \ref{tab:clinical}, the high-dimensional representation space ($d = 768$) of the diagnostic embedding confirms a brittle nature under continuous variance injection. When subjected to a noise variance of $\sigma^2 = 1.0$, the system undergoes  a significant collapse, resulting the terminal cosine similarity to drop to 0.4671. Illustrated in Figure \ref{fig:embedding_drift}, this nonlinear decay equates to a 53.29\% semantic degradation, suggesting that the multi-agent network has permanently lost the clinical context of the original ground-truth anchor. This isometric collapse proves that sequential, autoregressive prompting architectures are unable to survive unmitigated random drift in distributed environments without robust topological error correction.

\begin{table}[ht]
\centering
\caption{Terminal cosine similarity and semantic degradation of Bio\_ClinicalBERT [CLS] embeddings for $\sigma^2 \in \{0.1, 0.5, 1.0\}$.}
\small
\begin{tabular}{ccc}
\toprule
\textbf{Noise Variance ($\sigma^2$)} &
\textbf{Terminal Cosine Sim} &
\textbf{Semantic Degradation (\%)} \\
\midrule
0.1 & 0.8493 & 15.07\% \\
0.5 & 0.5939 & 40.61\% \\
1.0 & 0.4671 & 53.29\% \\
\bottomrule
\end{tabular}
\label{tab:clinical}
\end{table}

\subsection{Asymptotic Variance Trajectories and Systemic Instability}
\begin{figure}[ht]
\centering
\includegraphics[width=\linewidth]{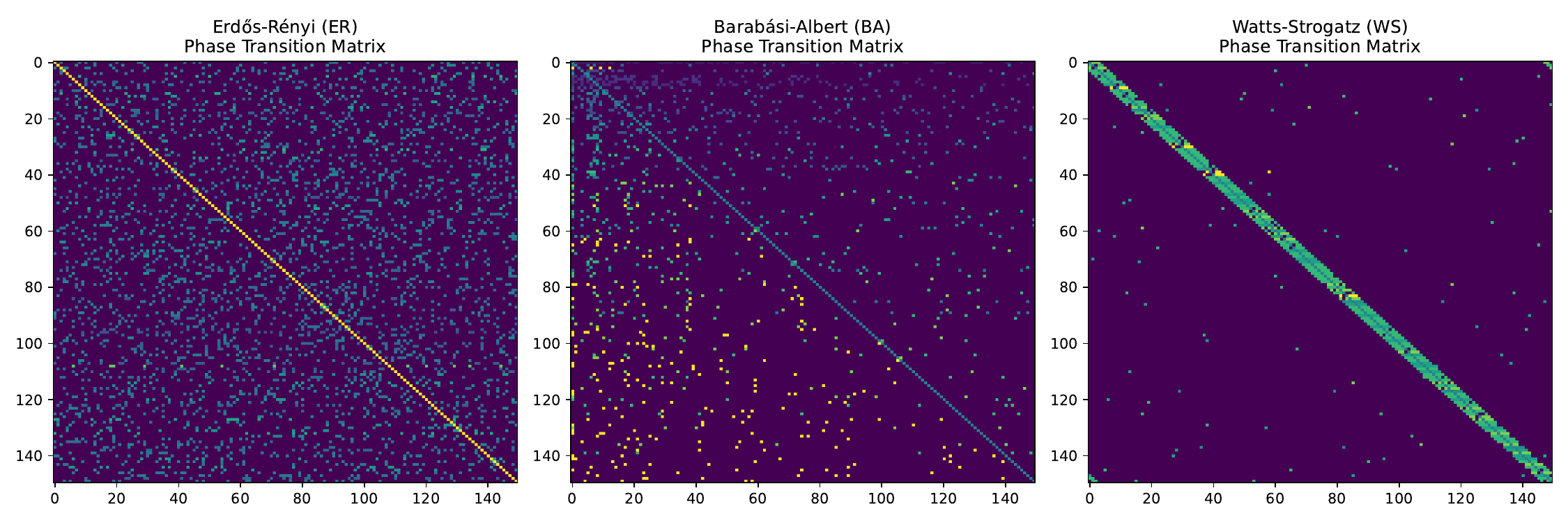}
\caption{System evolution matrices $M = (1-\gamma)I + \gamma W$ for Erdős-Rényi, Barabási-Albert, and Watts-Strogatz networks ($N = 150$, $\gamma = 0.90$).}
\label{fig:stage5_pair}
\end{figure}

The root of this semantic degradation is visually mapped in the system evolution matrices. Figure \ref{fig:stage5_pair} outlines the phase transition mapping across the three distinct architectures. The ER architecture shows that the transition probabilities are uniformly distributed, allowing random noise to disperse evenly across the continuous state space. However, the BA and WS architectures impose a sharp topological bottleneck; their spatial density constrains the transition probability mass to confine within densely connected local cliques, resulting in prevention of global state diffusion. This structural asymmetry guarantees systemic consensus failure, as formalized by the instability amplification ratio ($\rho$) in Table \ref{tab:stage5_trajectories}. The ER network mitigates the noise injection, yielding a minimal variance amplification ($\rho = 1.0766$). In contrast, the high clustering coefficient of the WS topology behaves as an echo chamber for hallucinated diagnostic data, amplifying the baseline variance by 51.81\% ($\rho = 1.5181$).

\begin{table}[ht]
\centering
\caption{Global state variance at $t = 0$, $t = 25$, and $t = 50$, with instability amplification ratios for Erdős-Rényi, Barabási-Albert, and Watts-Strogatz networks.}
\small
\begin{tabular}{ccccc}
\toprule
\textbf{Topology Architecture} &
\textbf{Initial ($t=0$)} &
\textbf{Midpoint ($t=25$)} &
\textbf{Peak ($t=50$)} &
\textbf{Amplification ($\rho$)} \\
\midrule
Erd\H{o}s--R\'enyi (ER) & 0.0101 & 0.0103 & 0.0109 & 1.0766 \\
Barab\'asi--Albert (BA) & 0.0101 & 0.0117 & 0.0124 & 1.2268 \\
Watts--Strogatz (WS) & 0.0101 & 0.0143 & 0.0153 & 1.5181 \\
\bottomrule
\end{tabular}
\label{tab:stage5_trajectories}
\end{table}

\begin{figure}[ht]
\centering
\includegraphics[width=\linewidth]{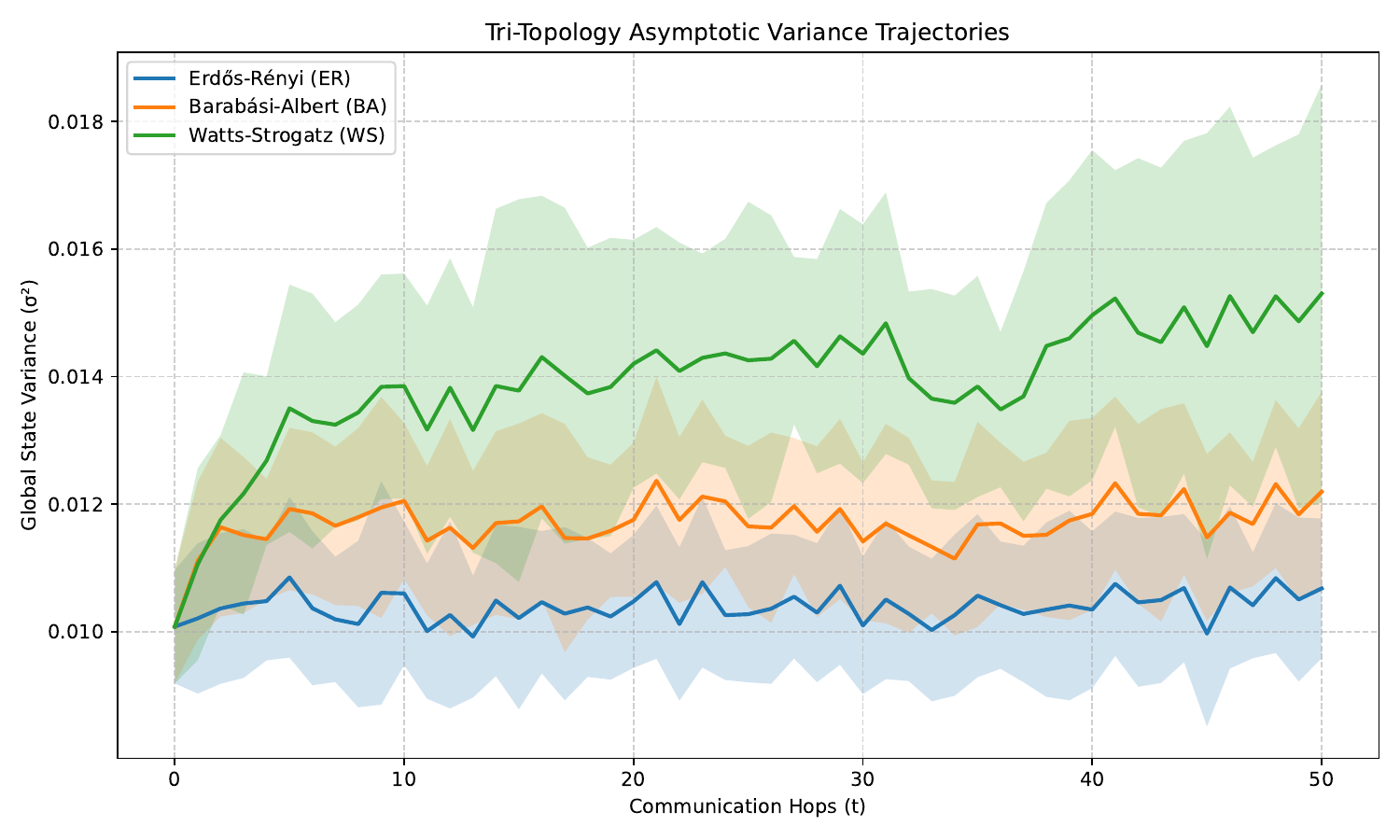}
\caption{Longitudinal global state variance ($\sigma^2$) trajectories over 50 communication hops for Erdős-Rényi, Barabási-Albert, and Watts-Strogatz architectures under a critical noise injection ($\sigma^2 = 0.01$, 20 Monte Carlo trials).}
\label{fig:stage5_p3}
\end{figure}

Moreover, the temporal evolution of the network confirms the inevitability of structural collapse. As illustrated in Figure \ref{fig:stage5_p3}, all three networks initiate from an identical baseline state variance ($t=0$). Whereas, the trajectories significantly diverge as the communication cascade continues. The ER network exhibits  asymptotic stability; the variance curve remains almost completely flat through the midpoint ($t=25$), totally absorbing the noise infusion. In contrast, the WS and BA topologies demonstrate a nonlinear acceleration in variance. The visibly widening confidence intervals in the WS trajectory reflect the compounding uncertainty spreading through its local sub-graphs. By the peak communication horizon ($t=50$), the small-world network has completely decoupled from the initial semantic anchor. This terminal trajectory confirms without the presence of ample algebraic connectivity to override matrix bottlenecks, the global state variance will cascade entirely, completely erasing the multi-agent diagnostic consensus.

\section{Discussion}
The results of our study challenge the conventional topological assumptions regarding multi-agent system design. By establishing rigorous mathematical metrics, we quantify the spectral vulnerability and variance trajectories of distinct network architectures. Furthermore, we demonstrate that structural bottlenecks are the primary catalyst for semantic degradation in iterative LLMs. The following subsections explicate these mathematical findings within the broader imperatives of clinical diagnostic safety and future architectural design.

\subsection{The Collapse of the Scale-Free Assumption}
The prevailing consensus in multi-agent system design posits that scale-free topologies, such as the Barabási-Albert (BA) architecture, provide superior robustness due to their highly interconnected hub nodes. Our phase transition matrices mathematically deconstruct this assumption for semantic communication. Rather than facilitating rapid information dispersal, structural bottlenecks in BA and Watts-Strogatz (WS) systems actively confine hallucinated data within dense local cliques. Without the capacity for rapid global diffusion, localized semantic errors compound autonomously. This clustering effect permanently prevents network-wide consensus, rendering hub-based architectures fundamentally non-viable for autonomous clinical reasoning cascades.

\subsection{Ramifications for Clinical NLP Diagnostics}
In the domain of clinical NLP, this architectural vulnerability translates into severe diagnostic safety risks. When a multi-agent network applying Bio\_ClinicalBERT embeddings approaches entropy saturation ($H_\infty \approx 5.947$), the system is not just generating an inaccurate output; it is confident in its error. The utilization of WS or BA architectures in medical diagnostic frameworks fundamentally guarantees that stochastic semantic drift will be intensified rather than mitigated. As we proved using the terminal cosine similarity degradation, the initial ground-truth is overwritten by hallucinated consensus. As a result, reliance on small-world or scale-free communication graphs for agent-based medical diagnostics induces an unacceptable threshold of systemic instability.

\subsection{Architectural Imperatives for Semantic Stability}

To ensure diagnostic safety, multi-agent architectures must algorithmically enforce uniformly distributed transition probabilities similar to the Erd\H{o}s-R\'enyi (ER) model, regardless of the physical or logical limitations of the underlying hardware. Engineering semantic stability needs maintaining a strict lower bound on algebraic connectivity ($\lambda_{2_{\min}}$) to guarantee that transition probability mass is not confined in local neighborhoods.

This algebraic threshold is an absolute prerequisite. Future architectures of clinical multi-agent frameworks must integrate dynamic topological error correction, as formalized in Algorithm \ref{alg:spectral_rewiring}. By actively monitoring the Fiedler value of the graph Laplacian at each communication hop ($t$), the system can independently inject randomized cross-clique edges whenever connectivity drops below the critical safety threshold ($\lambda_{2_{\min}}$).

\clearpage
\begin{algorithm}[H]
\caption{Dynamic Spectral Gap Monitoring and Targeted Cross-Partition Rewiring}
\label{alg:spectral_rewiring}
\begin{algorithmic}[1]
\Require Critical Algebraic Threshold $\lambda_{2_{\min}}$, Initial Topology $G_t=(V, E_t)$
\Ensure Topologically Corrected Graph $G_{t+1}$

\State Construct Laplacian Matrix $L_t \gets D_t - A_t$
\State Execute Eigen-decomposition: $L_t v_k = \lambda_k v_k$
\State Extract Algebraic Connectivity $\lambda_2(L_t)$ and Fiedler Vector $v_2 \in \mathbb{R}^N$

\If{$\lambda_2(L_t) < \lambda_{2_{\min}}$}
    \State Initialize added edge set $E_{add} \gets \emptyset$
    \While{$\lambda_2(L_t) < \lambda_{2_{\min}}$}
        \State Compute spectral perturbation gradient for candidate edges:
        \Statex \hspace{\algorithmicindent}$\Delta \lambda_{2, (i,j)} \approx (v_2(i) - v_2(j))^2$
        \State Identify optimal cross-partition boundary nodes:
        \Statex \hspace{\algorithmicindent}$u \gets \arg\max_i (v_2(i))$
        \Statex \hspace{\algorithmicindent}$v \gets \arg\min_j (v_2(j))$
        \If{$(u,v) \notin E_t \cup E_{add}$}
            \State $E_{add} \gets E_{add} \cup \{(u,v)\}$
            \State Update Adjacency: $A_t(u,v) \gets 1, \quad A_t(v,u) \gets 1$
            \State Recompute Laplacian: $L_t \gets \text{diag}(\sum_j A_t(\cdot, j)) - A_t$
            \State Update Fiedler value $\lambda_2(L_t)$ via localized Rayleigh quotient update
        \Else
            \State Sample candidate pair $(u,v) \sim \text{Uniform}(V \times V \setminus E_t)$ as stochastic fallback
        \EndIf
    \EndWhile
    \State $G_{t+1} \gets (V, E_t \cup E_{add})$
\Else
    \State $G_{t+1} \gets G_t$
\EndIf
\State \Return $G_{t+1}$
\end{algorithmic}
\end{algorithm}

\subsection{Limitations and Computational Overhead}
First, while computing exact eigen-decompositions requires $\mathcal{O}(N^3)$ complexity, practical online deployments can reduce this overhead to $\mathcal{O}(\lvert E \rvert k)$ using sparse iterative Lanczos or Krylov subspace methods. Second, our isotropic noise model $\zeta_i \sim \mathcal{N}(0, x_i(T)I)$ serves as an analytical upper bound on semantic drift; real-world LLM errors exhibit anisotropic drift along specific directional sub-manifolds of the transformer latent space. Future work will extend this framework to directionally constrained covariance matrices.

Furthermore, our mathematical architecture depends on simplifying abstractions---mapping high-dimensional representations to a scalar belief state using cosine similarity against an offline reference anchor, and matching semantic drift through additive Gaussian noise perturbations. While tractable for baseline asymptotic analysis, these approximations are not able to fully capture content-conditioned, systemic LLM error modes or multi-turn conversational dynamics.

Finally, the baseline established in this study uses a homogeneous multi-agent architecture utilizing uniformly dimensional Bio\_ClinicalBERT embeddings. The premise of isotropic semantic drift may not hold up in heterogeneous systems where diagnostic processes are dispersed across various frameworks, such as using text-based NLP agents with computer vision models processing radiological scans. Handling such hallucinations across diverse embedding manifolds is still a key frontier for guaranteeing the robustness of clinical multi-agent frameworks.

\section{Conclusion}
The utilization of iterative LLMs within multi-agent diagnostic frameworks demands a fundamental departure from conventional topological assumptions. Our study established that hub-centric and highly clustered network architectures are structurally incompatible with reliable semantic communication. By quantifying the spectral vulnerability of these networks, we proved that localized bottlenecks confine and compound hallucinated data. To ensure clinical diagnostic safety, future research must prioritize global state diffusion over localized efficiency. The established implementation of dynamic spectral monitoring, which imposes a strict lower bound on algebraic connectivity ($\lambda_{2_{\min}}$), provides a mathematically rigorous mechanism to prevent semantic collapse. Finally, securing the reliability of agent-based clinical reasoning necessitates treating framework stability as a non-negotiable imperative in multi-agent system design.

\section*{Code Availability and Reproducibility}
To ensure complete computational reproducibility of all analytical results, structural simulations, and visual plots presented in this study, the complete source code has been made publicly available in a dedicated GitHub repository at:\\
\url{https://github.com/amribanerjee/spectral-semantic-cascade}

The code is provided as a structured Jupyter Notebook named \texttt{script.ipynb}, organized into independent execution stages matching the evaluation pipeline:
\begin{itemize}
    \item \textbf{Stage 1}: Spectral topology initialization, graph Laplacian computation, and Fiedler vector distribution generation.
    \item \textbf{Stage 2}: Information-theoretic entropy saturation modeling and asymptotic Kullback-Leibler divergence trajectories.
    \item \textbf{Stage 3}: Multi-agent consensus cascade simulation and global state variance tracking.
    \item \textbf{Stage 4}: Bio\_ClinicalBERT latent space degradation and embedding cosine similarity decay under variance injection.
    \item \textbf{Stage 5}: Tri-topology phase transition matrices, eigenvector mass concentrations, and multi-trial statistical trajectories with confidence bands.
\end{itemize}

All simulations impose a strict, fixed random seed (\texttt{seed = 42}) to ensure deterministic output generation across independent execution environments. Complete package dependencies, hardware configuration notes, and step-by-step execution instructions are documented within the repository's README file.

\section*{Acknowledgements}

The author would like to acknowledge Zoravar Singh for his contributions to the literature review and the structuring of the related works section, which provided necessary contextual foundation for the topological analysis presented in this manuscript.

\section*{Declarations}

\textbf{Use of AI-Assisted Technologies in Writing:} During the preparation of this manuscript, the author utilized a large language model to polish and enhance the readability of the text. After using this tool, the author thoroughly reviewed and edited the content, and assumes full responsibility for the final publication's mathematical and theoretical integrity.

\bibliographystyle{unsrt}
\bibliography{references}

\end{document}